\documentclass{iau} 
\usepackage{graphicx}
\title[Noise, friction and ROI]{Noise, friction and the radial-orbit instability in anisotropic stellar systems: stochastic $N-$body simulations}
\author[P.\ Di Cintio \& L.\ Casetti]{Pierfrancesco Di Cintio$^{1,2,3}$ \and Lapo Casetti$^{2,3,4}$}
\affiliation{$^1$Enrico Fermi Research Center (CREF), Via Panisperna 89A,
I-00184, Rome, Italy \\ email: {\tt pierfrancesco.dicintio@unifi.it} \\[\affilskip]
$^2$INFN, Sezione di Firenze, via G.\ Sansone 1,
I-50019, Sesto Fiorentino (FI), Italy\\
$^3$Dipartimento di Fisica e Astronomia, Universit\`a di Firenze,\\ via G.\ Sansone 1,
I-50019, Sesto Fiorentino (FI), Italy\\
$^4$INAF-Osservatorio astrofisico di Arcetri, largo E.\ Fermi 5, I-50125, Firenze, Italy}
\pubyear{2021}
\volume{364}  
\jname{Multiscale (time and mass) dynamics of space objects}
\editors{A. Celletti, C. Beaug\'e, C. Gales \& A. Lem\^aitre, eds.}
\begin{document}
\maketitle
\begin{abstract}
By means of numerical simulations we study the radial-orbit instability in anisotropic self-gravitating $N-$body systems under the effect of noise. We find that the presence of additive or multiplicative noise has a different effect on the onset of the instability, depending on the initial value of the orbital anisotropy.
\end{abstract}
\firstsection 
\section{Introduction}
Spherically symmetric, self-gravitating collisionless equilibrium systems with a large fraction of the kinetic energy stored in low angular momentum orbits 
are known to be dynamically unstable. The associated instability is known as Radial Orbit Instability (hereafter ROI, see e.g. \cite{poly15} and references therein). Usually, the amount of radial anisotropy in a spherical system is quantified by introducing the Fridman-Polyachenko-Shukhman parameter (see \cite{bt08})
\begin{equation}
 \xi\equiv\frac{2K_r}{K_t},
\end{equation}
where the radial and tangential kinetic energies are given respectively by
\begin{equation}
K_r=2\pi\int\rho(r)\sigma^2_r(r)r^2{\rm d}r,\quad K_t=2\pi\int\rho(r)\sigma^2_t(r)r^2{\rm d}r,
\end{equation}
 $\rho$ is the system density, and $\sigma^2_r$ and $\sigma^2_t$ are the radial and tangential phase-space averaged square velocity components, respectively. For isotropic systems $\xi=1$. Numerical simulations show that the ROI typically occurs for $\xi\gtrsim1.7$, even though it is well known that the "real" critical value of $\xi$ above which the given system is unstable, depends on the specific phase-space structure of the initial condition under consideration.\\
\indent The ROI it is frequently invoked as the mechanism responsible for the triaxiality of the elliptical galaxies and the formation of bars in disk galaxies. However, little is known on the effective nature of the underlying mechanism or its near- or far-field origin (see e.g. \cite{poly15,dcn1} and references therein). Recently, \cite{perez10} introduced a novel interpretation of ROI as a (effective) dissipation-induced phenomenon. In this preliminary work we investigate their argument by means of direct $N-$body simulations with a controllable source of (external) noise and dissipation.  
\begin{figure}
  \includegraphics[width=0.8\textwidth]{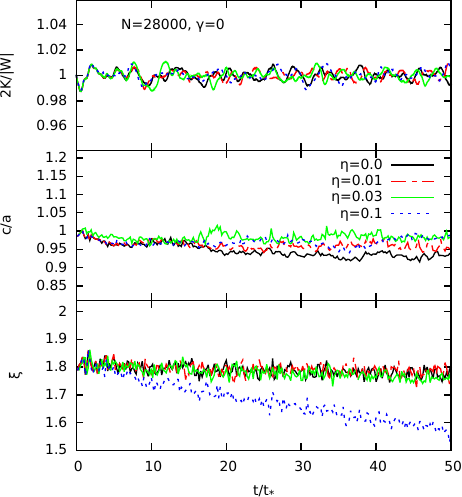}
\caption{For an initially mildly anisotropic ($\xi_0=1.8$) flat-cored model ($\gamma=0$) with $N=28000$, evolution of the virial ratio $2K/|W|$ (top panel), axial ratio $c/a$ (middle panel) and anisotropy parameter $\xi$ (bottom panel).}
\label{figvir}       
\end{figure}
\section{Methods}
We study the stability of a family of $\gamma-$models with density profile given by 
\begin{equation}\label{dehnen}
\rho(r)=\frac{3-\gamma}{4\pi}\frac{Mr_c}{r^\gamma(r+r_c)^{4-\gamma}},
\end{equation}
with total mass $M$, scale radius $r_c$ and logarithmic density slope $\gamma$. In order to generate the velocities for the simulation particles we use the standard rejection technique to sample the anisotropic equilibrium phase-space distribution function $f(Q)$, obtained for a given (spherical) density-potential couple $(\rho,\Phi)$ linked by the Poisson equation $\Delta\Phi=4\pi G\rho$, applying the usual Osipkov-Merritt reparametrization (\cite{osipkov85,merritt85}) of the \cite{edd16} integral inversion
\begin{equation}\label{OM}
f(Q)=\frac{1}{\sqrt{8}\pi^2}\int_Q^{0}\frac{{\rm d}^2\rho_a}{{\rm d}\Phi^2}\frac{{\rm d}\Phi}{\sqrt{\Phi-Q}}.
\end{equation}
In Equation (\ref{OM}) 
\begin{equation}
Q=E+{J^2}/{2r_a^2}, 
\end{equation}
 and $E$ and $J$ are the particle's energy and angular momentum per unit mass, respectively. The quantity  $r_a$ is the so-called anisotropy radius, and $\rho_a$ the augmented density, defined by 
 \begin{equation}
 \rho_a(r)\equiv\left(1+{r^2}/{r_a^2}\right)\rho(r).
 \end{equation}
 For our specific choice of $\rho(r)$ in Eq. (\ref{dehnen}), the model's potential is given by
 \begin{eqnarray}
 \Phi(r)&=&-\frac{GM}{(2-\gamma)r_c}\left[1-\left(\frac{r}{r+r_c}\right)^{2-\gamma}\right]\quad {\rm for}\quad\gamma\neq 2;\nonumber \\
 \Phi(r)&=&\frac{GM}{(2-\gamma)r_c}\ln\frac{r}{r+r_c} \quad {\rm for}\quad\gamma=2.
 \end{eqnarray}
 The anisotropy radius $r_a$ controls the extent of anisotropy of the model so that,  the velocity-dispersion tensor is nearly isotropic for $r<r_a$, and increasingly  radially anisotropic for $r>r_a$, thus small values of $r_a$ are associated to more radially anisotropic systems, i.e. larger values of $\xi$.\\
\indent Throughout this work we assume units such that $G=M=r_c=1$, so that the dynamical time and the scale velocity become $t_*=\sqrt{r_c^3/GM}$ and  $v_*=r_c/t_*$ and are both equal to unity. Individual particle masses are therefore $m=1/N$.\\ 
\begin{figure}
  \includegraphics[width=0.8\textwidth]{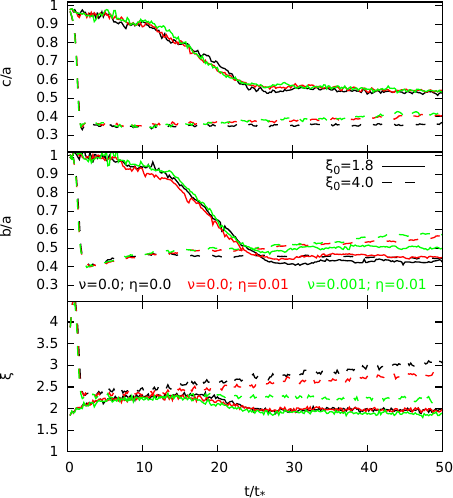}
\caption{For initially mildly anisotropic ($\xi_0=1.8$, solid lines) and strongly anisotropic ($\xi_0=4$, dashed lines) Hernquist models ($\gamma=1$): evolution of the axial ratios $c/a$ (top panel), $b/a$ (middle panel) and anisotropy parameter $\xi$ (bottom panel).}
\label{figca}       
\end{figure}
\indent In order to consider the effect of noise and dissipation, we express the particles' dynamics in terms of Langevin-like equations (e.g. see \cite{1980PhR....63....1K}) of the form 
\begin{equation}\label{langeq}
\ddot{\mathbf{r}}_i=-\nabla\Phi(\mathbf{r}_i)-\nu\mathbf{v}_i+\mathbf{F}(\mathbf{r}_i),
\end{equation}
where, in our case, the acceleration $-\nabla\Phi$ on each particle is evaluated  self-consistently by direct sum over all other particles, $\nu$ is the dynamical friction [\cite{1943ApJ....97..255C,1949RvMP...21..383C}] coefficient, and $\mathbf{F}(\mathbf{r})$ a fluctuating force (per unit mass).\\
\indent In our numerical simulations we solved Eqs.(\ref{langeq}) with the so-called quasi-symplectic \cite{2004PhRvE..69d1107M} scheme with fixed time-step $\Delta t=10^{-2}t_*$, in the same fashion as \cite{pasquato20} and \cite{dcn2}.\\
\indent We note that, a similar approach could also be extended to the study of protoplanetary disks in dense environments under the effect of flyby stars, since in principle the disk hydrodynamics and the stellar dynamics have different time scales in a numerical simulation so the effect of passing stars could be simplified as a stochastic process (e.g. see \cite{capuzzo}). 
\section{Numerical simulations and discussion}
Following \cite{kandrup99}, \cite{terzic03} and \cite{kandrup04b}, we have implemented three different forms of noise: {\it i)} additive noise without friction (i.e. $\nu=0$ in Eq. \ref{langeq}). {\it ii)} additive noise connected to friction via the Fluctuation-Dissipation Theorem such that 
\begin{equation}
\eta^2=\Theta\nu/t_c,
\end{equation}
\begin{figure}
  \includegraphics[width=0.8\textwidth]{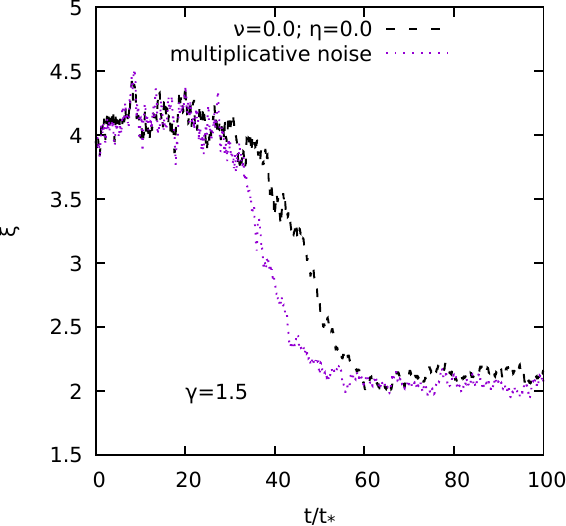}
\caption{Evolution of the anisotropy parameter $\xi$ for a system starting with $\gamma=1.5$ and $\xi_0=4$}
\label{xi}       
\end{figure}
where $\eta$ is the typical amplitude of the Gaussian distributed force $\mathbf{F}$, $\Theta$ is the system's temperature (proportional to the velocity dispersion $\sigma$) and $t_c$ is the autocorrelation time of the noise. {\it iii)} multiplicative noise with friction where the dynamical friction coeffcient is explicitly dependent on the particle velocity as 
\begin{equation}\label{nu}
\nu=4\pi G^2\rho_*(m+m_*)\ln\Lambda\frac{\Psi(v)}{v^3},
\end{equation}
where $G$ is the gravitational constant, $\rho_*$ is the mass density of a (fictitious) background of particles of mass $m_*$, $\ln\Lambda$ is the Coulomb logarithm, $v=||\mathbf{v}||$, and
\begin{equation}\label{velfunct}
\Psi(v)=4\pi\int_0^{v}f(v_*)v_*^2{\rm d}v_*,
\end{equation}
is the fractional velocity volume function (see e.g. \cite{bt08}).\\ 
\indent In Figure \ref{figvir} we show the evolution of the virial ratio $2K/|W|$, where $K$ is the total kinetic energy and $W=\sum_N m_i\mathbf{r}_i\cdot\nabla\Phi(\mathbf{r}_i)$ the virial function; the minimum to maximum axial ratio $c/a$ and the anisotropy parameter $\xi$ for an initially mildly anisotropic $\gamma=0$ system with $\xi_0\simeq 1.8$, subjected to frictionless noise for various values of noise amplitude $\eta$. In all cases, the presence of the additive noise does not take the system out of virial equilibrium, while for low values of $\eta$ (and also large values of $t_c$, not shown here) some deviations from the spherical symmetry are evident. In general, large values of $\eta$ have a somewhat stabilizing effect against ROI, as less and less deviations from $c/a=1$ are observable and $\xi$ tends to decrease for $\eta>0.05$.\\
\indent In Figure \ref{figca} we compare the evolution of the axial ratios $c/a$ and $b/a$ and the anisotropy parameters for $\gamma=1$ models with $\xi_0=4$ and 1.8 for additive noise with and without friction. In general, the presence of noise or noise plus friction does not have a significant effect on the onset of ROI for models with a steeper cusp (generally more unstable, as they admit a larger degree of wildly chaotic orbits see \cite{dcs,dcs2}) and low values of the initial anisotropy (i.e. $\xi_0=1.8$), while for larger values of $\xi_0$, corresponding to a more violent instability, the evolution of the triaxiality and the anisotropy are affected by the presence of noise, with systematically less anisotropic and more "triaxial" end states associated to the presence of larger amounts of noise and friction. Introducing a multiplicative noise (with velocity dependent friction coefficient) complicates the picture even further with as it apparently it does not alter significantly the evolution of the axial ratios, nor the final values attained by $\xi$ even for extremely anisotropic models with steep cusps, while it seems to somewhat anticipate the time at which the anisotropy parameter starts moving to lower values (i.e. unstable models become more isotropic earlier), as shown in Figure \ref{xi} for a system with $\gamma=1.5$.\\
\indent From these preliminary results we speculate that the mechanisms leading to ROI might work differently in real systems subjected to different form of internal or environment-related sources of noise, as well as different central concentrations (see e.g. \cite{trenti06}). In particular, we speculate that multicomponent systems with different anisotropy profiles for each component could develop the ROI in a substantially different fashion as their single component counterparts. We will explore this matter further in a forthcoming publication (\cite{dzc}), studying the stability of a family of \cite{zocchi15} models with more mass components with tunable degree of radial anisotropy.

\end{document}